\documentclass[11pt]{article}
\usepackage{moriond,epsfig,float}
\usepackage{sidecap}

\def\Journal#1#2#3#4{{#1} {\bf #2}, #3 (#4)}

\def\PRD{{\em Phys. Rev.} D}

\def\Nat{{\em Nature}}
\def\Apjs{{\em ApJ Suppl.}}

\def\be{\begin{equation}}
\def\ee{\end{equation}}
\def\bea{\begin{eqnarray}}
\def\eea{\end{eqnarray}}

\begin{document}
\vspace*{4cm}
\title{CLOVER -- A NEW INSTRUMENT FOR MEASURING THE $B$-MODE POLARIZATION
  OF THE CMB}

\author{ A.C. Taylor, A. Challinor, D. Goldie, K. Grainge,
  M.E. Jones, A.N. Lasenby, S. Withington and G. Yassin  }

\address{Astrophysics Group, Cavendish Laboratory, Madingley Road,
  Cambridge, CB3 0HE, England}

\myauthor{ W.K. Gear, L. Piccirillo, P. Ade, P.D. Mauskopf, B. Maffei
  and G. Pisano  }

\myaddress{Dept. of Physics and Astronomy, University of Wales,
  Cardiff, CF24 3YB,Wales}

\maketitle\abstracts{We describe the design and expected performance
  of Clover, a new instrument designed to measure the $B$-mode
  polarization of the cosmic microwave background. The proposed
  instrument will comprise three independent telescopes operating at
  90, 150 and 220~GHz and is planned to be sited at Dome C,
  Antarctica.  Each telescope will feed a focal plane array of 128
  background-limited detectors and will measure polarized signals over
  angular multipoles 20\textless $\ell$ \textless1000.  The unique
  design of the telescope and careful control of systematics should
  enable the $B$-mode signature of gravitational waves to be measured
  to a lensing-confusion-limited tensor-to-scalar ratio $r\sim0.005$.}

\section{Introduction}

In recent years the power spectrum of temperature anisotropies in the
CMB has been measured with ever increasing precision and resolution
allowing tighter constraints to be placed on cosmological models and
parameters.  However, yet more information can be obtained via
measurements of the polarization of the CMB. Thomson scattering of the
CMB at recombination and reionization gives rise to linear
polarization which can be decomposed into a curl-free part ($E$-mode
polarization) and a divergence-free part ($B$-mode polarization).  The
primordial scalar perturbations that are responsible for the observed
temperature anisotropies can produce only $E$-mode polarization in
linear theory but a cosmological background of gravitational waves
(tensor modes), such as that generated in most models of inflation,
produces both $E$- and $B$-mode polarization. Detection of either of
the $E$- or $B$-mode polarization provides a wealth of information
complementary to that contained in the temperature anisotropies, and a
detection of the tensor contribution will provide a uniquely powerful
test of inflationary models. The predicted amplitudes of the $E$- and
$B$-mode polarization fluctuations are extremely small ($\sim$
5~$\mu$K and \textless 0.1~$\mu$K respectively) and hence measurement
of either requires both superb sensitivity and an unprecedented
control of systematics.  A detection of the $E$-mode polarization was
first reported by the DASI \cite{DASI} team and a measurement of the
temperature-polarization cross-correlation on larger angular scales
($\ell < 200$) has been published by the WMAP \cite{WMAP} team.
However, as yet there has been no detection of the $B$-mode
polarization of the CMB.  Here we describe a new experiment, Clover,
which has been designed with the sensitivity and control of
systematics required to allow a detection of the gravitational wave
contribution to $B$-mode polarization.

\section{Instrument overview}

The main science goal of Clover is to measure the power spectrum of
the $B$-mode polarization in the multipole range $\ell$ = 20--1000,
with the aim of making measurements down to a sensitivity limited by
the contamination due to foreground lensing of the $E$-mode signal for
multipoles $\ell <$ 200. Clover consists of three completely
independent telescopes, operating at 90, 150 and 220~GHz and hence
will allow spectral subtraction of foreground components. Each
telescope consists of four separate, co-pointed optical assemblies
(Fig.~\ref{fig:optics}, left), each fed by an 8$\times$8 array of feed
horns and scaled in size such that each has a beam on the sky of 15 arcmin.
The signal from each horn is separated into the two independent
polarization states, converted to circular polarization, phase
modulated and then correlated (Fig.~\ref{fig:optics}, right).  The two
correlator outputs encode the Stokes parameters $I$, $Q$ and $U$.  The
outputs from each corresponding pixel in the four optical assemblies
are then summed incoherently before being detected by a TES bolometer.
There are thus 256 horns per telescope but only 64 simultaneously
observed pixels, since the optical assemblies are co-pointed.  The
sensitivity, however, is equivalent to 256 individually-detected
pixels. Stokes parameters $Q$ and $U$ are measured instantaneously by
the phase modulation in the polarimeter, while $I$ is measured by
scanning the telescope. Table~\ref{tab:params} summarises the current
specification for Clover.

\begin{table}[t]
\caption{Specification summary for Clover.\label{tab:params}}
\begin{center}
\begin{tabular}{lccc}
\hline
Telescope frequency  & 90~GHz             & 150~GHz           & 220~GHz  \\
Bandwidth & 30~GHz     & 45~GHz                  & 60~GHz          \\
Pixel NET & 170 $\mu$Ks$^{1/2}$ & 215 $\mu$Ks$^{1/2}$ & 455 $\mu$Ks$^{1/2}$ \\
Array NET & 10.5 $\mu$Ks$^{1/2}$ & 13.4 $\mu$Ks$^{1/2}$ & 28.5 $\mu$Ks$^{1/2}$\\
Beam FWHM & 15 arcmin          &   15 arcmin       & 15 arcmin\\

\hline
\end{tabular}
\end{center}
\end{table}

\subsection{Telescope and mount}
\begin{figure}[t]
\includegraphics[width=4.6cm]{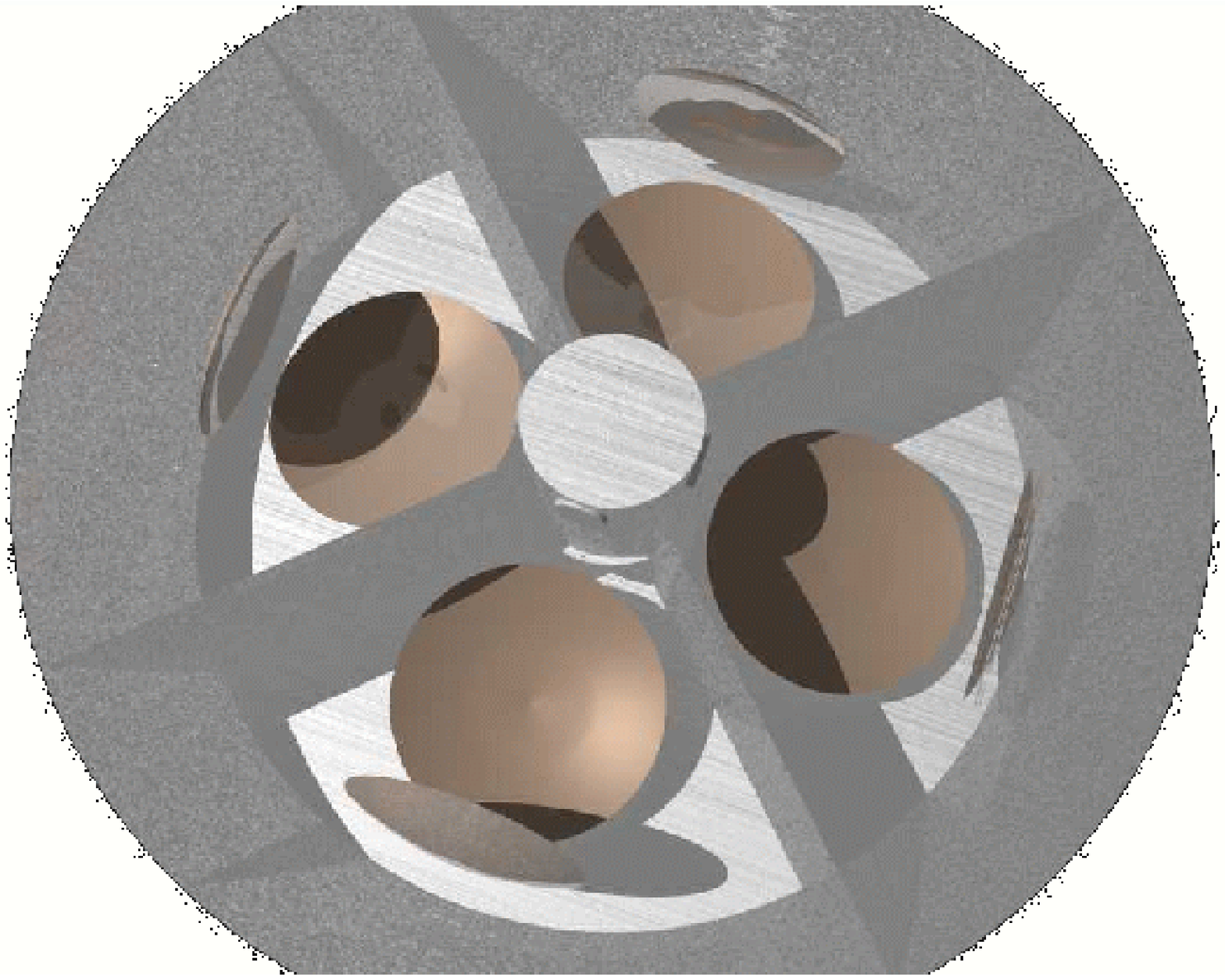}
\hspace{1cm}%
\includegraphics[width=4.6cm]{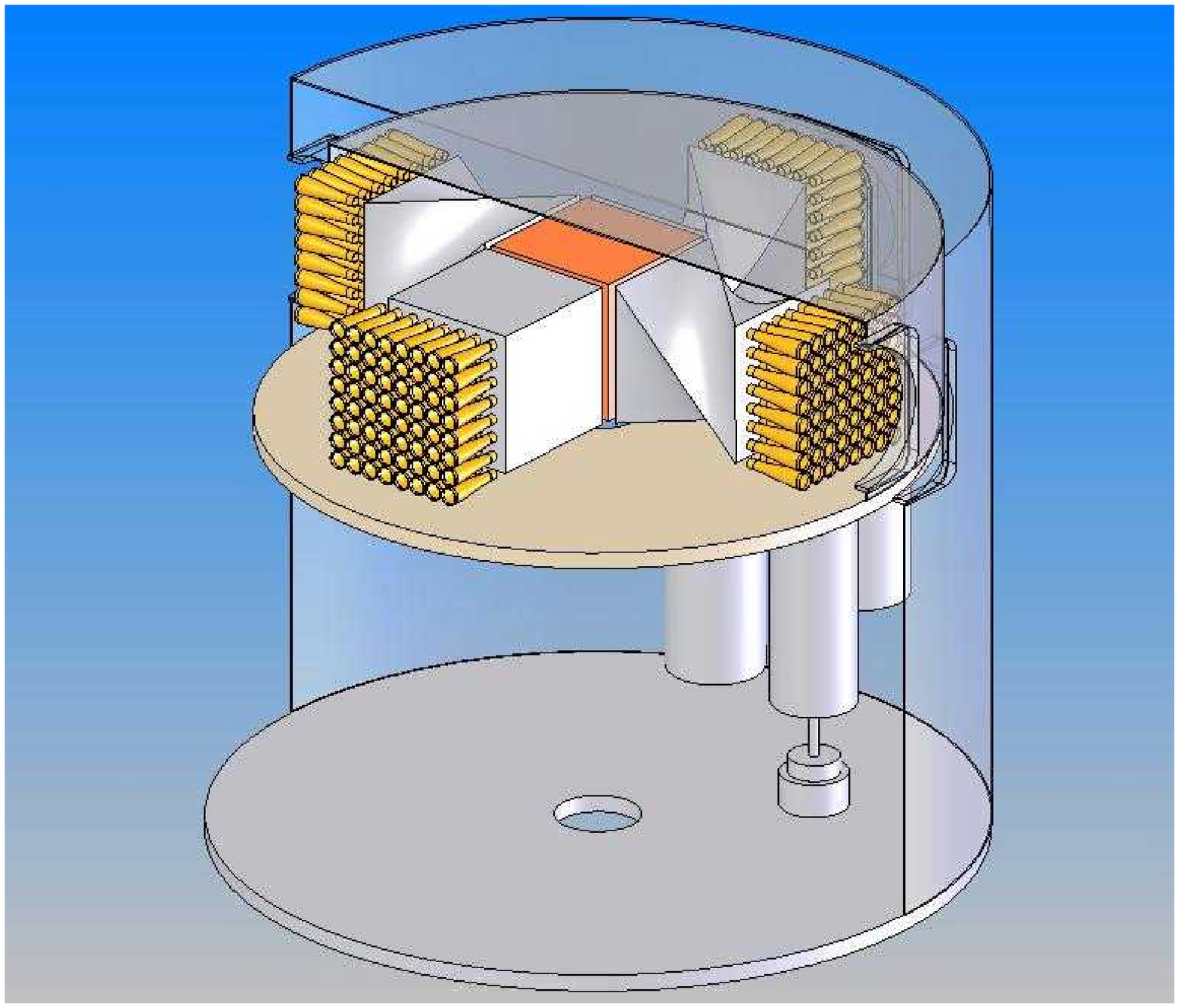}%
\hspace{2cm}%
\includegraphics[height=4.6cm]{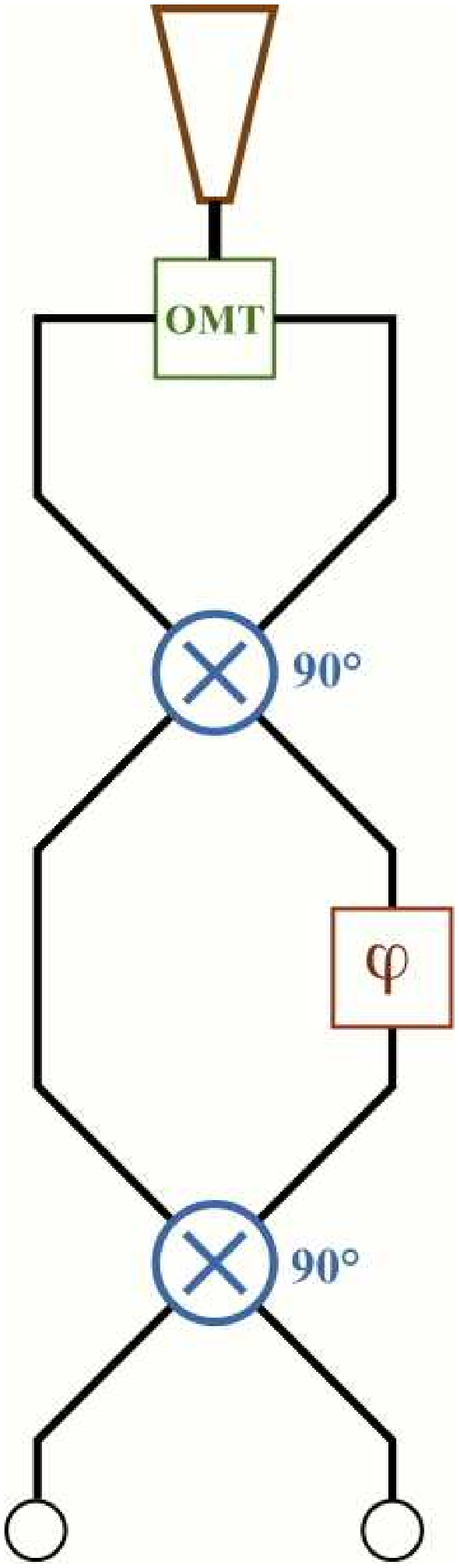}%
\caption{{\bf Left:} Schematic view of a single observing platform with four
  co-aligned telescopes each feeding an array of 64 feed-horns. {\bf
    Middle:} The central cryostat containing four 64-element feed
  arrays and the sections behind containing the hybrids and phase
  switches. Note the twisted waveguide sections mating the horn arrays
  to the detector array (central cube), allowing the same sky pixels
  to be brought to the same detector. {\bf Right:} Schematic of the
  pseudo-correlation receiver for a single feed.
\label{fig:optics}}
\end{figure}
The optical assembly of Clover comprises a Compact Range Antenna (CRA)
which is an offset-fed design exhibiting low off-axis beam distortion
and cross-polarization (\textless$-$35dB for an edge pixel).  The four
optical assemblies on each telescope are built around a single
cryostat which houses all four horn/polarimeter arrays and the
detector array (Fig.~\ref{fig:optics}, middle), and are mounted on a
common mount which allows altitude-azimuth tracking as well as
rotation of the entire optical structure around the pointing axis.
This arrangement allows the total optical throughput to be increased
relative to a single telescope design with the same size of focal
plane array, while the rotational symmetry allows physically different
arrays to be rotated into the same orientation thus providing
cross-checks on telescope-dependent systematics.

\subsection{Array and polarimeter}

The signals from each optical assembly are fed into an 8$\times$8 array of
corrugated feed-horns and then in waveguide to a
pseudo-correlation polarimeter (Fig.~\ref{fig:optics}, right). Each
polarimeter has two outputs, $D_{1}$ and $D_{2}$, given by

\begin{equation}
\begin{array}{rcl}
D_{1} & = & I + Q\cos\phi + U\sin\phi \\
D_{2} & = & I - Q\cos\phi - U\sin\phi,  \\
\end{array}\label{eq:spa}
\end{equation}

\noindent where $\phi$ is the phase introduced by a phase
modulator.  The detector outputs can thus clearly be used to determine
both $Q$ and $U$ at the sky pixel by taking the difference of the
detector outputs and phase-sensitively detecting.  The effect is
similar to that of a rotating wave plate, but is achieved without any
moving optics.  The intensity $I$ of the pixel is also obtainable from
the sum of the detector outputs but is not modulated.  Modulation of
the intensity is achieved by scanning of the array across the sky.

\subsection{Detector array}

The outputs from the pseudo-correlation receivers are detected using
an array of voltage-biased Transition Edge Sensors~\cite{louisa,ghassan} (TESs) operating at
300~mK.  Since the four optical assemblies feeding the horn-arrays are
co-pointed, the signals from common pixels in the four arrays can be
detected by the same TES. Each telescope has four arrays, containing
64 horns and 128 outputs (for the two polarizations), and so a total
of 128 TES detectors is required for each of the three Clover
telescopes.  Readout of the TES detectors is achieved by frequency
multiplexing of eight TES detectors onto one SQUID.

\section{Site and observing strategy}

It is intended that Clover will be installed at Dome C on the
Antarctic Plateau at an altitude of 3200~m. This site is one of the
premier locations for mm and sub-mm observations, providing dry and
atmospherically stable conditions which are comparable to, if not
slightly better than, the South Pole. In the first two years of
operation we aim to observe a connected region of sky of a few hundred
square degrees.  We plan to implement a multi-cross scanning strategy
consisting of observing a given patch of sky scanning over a fixed
azimuth range while keeping the elevation constant for a 2--3-hour
period. Once this time has elapsed, the centre azimuth and elevation
are changed to follow the sky patch and constant elevation scans are
repeated. This strategy results in a well cross-linked coverage of a
single sky area. The telescopes will also be periodically rotated
about the pointing axis to calibrate out instrumental effects and
improve the density and cross-linking of the sky coverage.

\section{Science predictions}

\begin{SCfigure}[0.9]
\psfig{figure=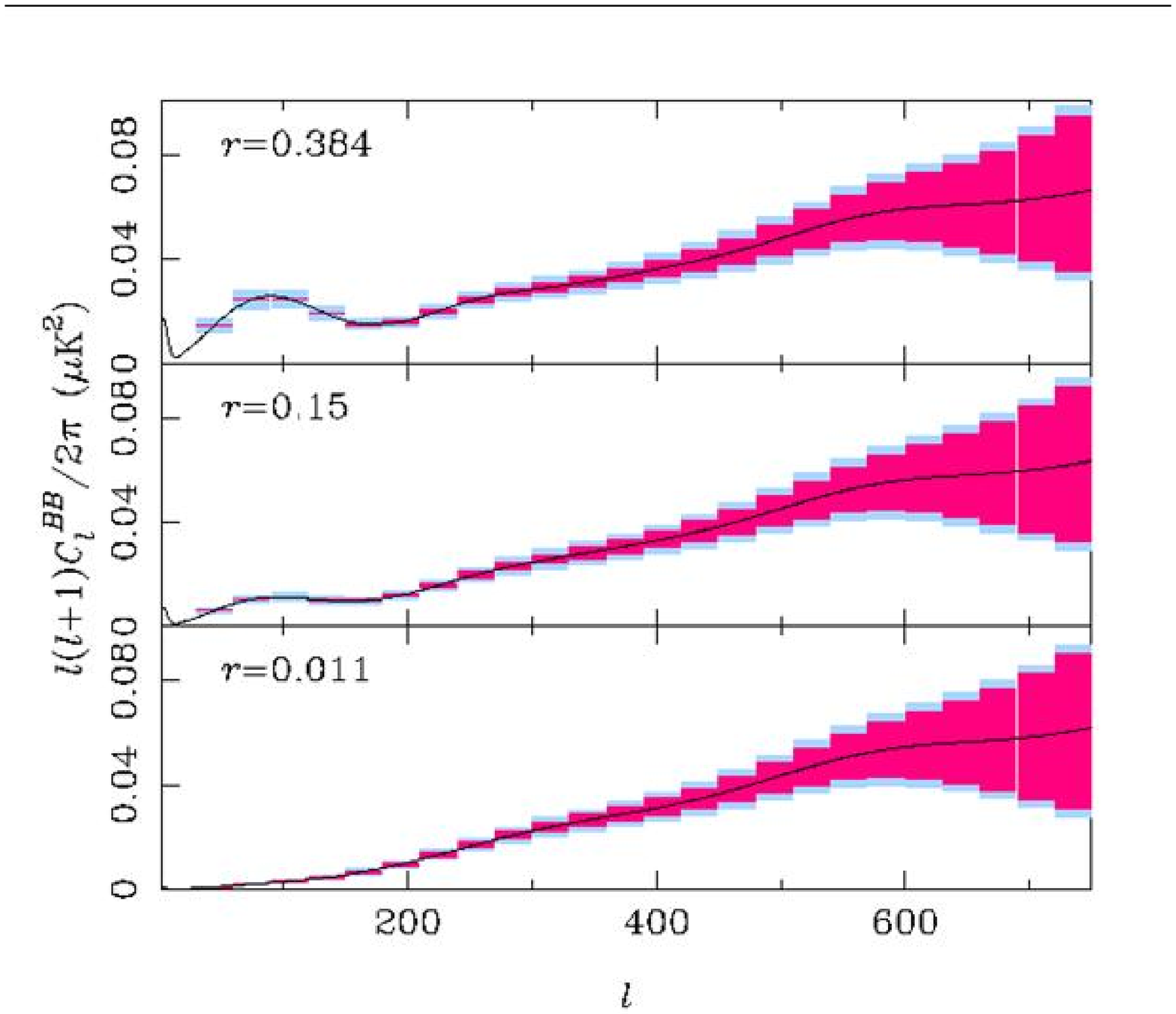,width = 7cm,clip}
\caption{Expected errors from Clover on the $B$-mode power spectrum for
  flat, $\rm {\Lambda CDM}$ inflation models. The upper panel has
  tensor-to-scalar ratio $r$=0.384 corresponding to the current 68-per
  cent upper limit from CMB and large-scale structure data$^{3}$.
  The middle panel is for $\phi^{2}$ inflation, with
  $r=0.15$. The lower panel is for small-field, parabolic inflation
  $[V\propto 1-(\phi/\phi_{*})^{2}]$ with $r$=0.011.  The inner error
  boxes are the contribution from instrument noise after foreground
  removal; the outer boxes also include sample variance from the
  gravitational wave signal and weak lensing.
\label{fig:predictions}}
\end{SCfigure}

Modelling of foreground removal with reasonable assumptions about the
level of Galactic contaminants suggests that the effective sensitivity
of the full instrument will be similar to that obtained by a single
frequency channel in the absence of foregrounds. From a two-year
experiment observing a near-circular survey region of radius 15
degrees we expect a thermal noise level after subtraction of
foregrounds of approximately 0.24~$\mu K$ per resolution element
(15-arcmin by 15-arcmin) to the Stokes parameters $Q$ and $U$. For
comparison, the expected r.m.s. of $Q$ and $U$ is 2.1~$\mu K$ at
15-arcmin resolution; 0.1~$\mu K$ of this arises from the $B$-mode
polarization generated by lensing, and 0.3$\sqrt r\,\mu K$ from
gravitational waves. The performance of Clover in measuring the total
$B$-mode power spectrum for three different values of the
tensor-to-scalar ratio is shown in Figure~\ref{fig:predictions} .  We
find that the one-sigma error on $r$, computed from the errors on
$C_{\ell}^{B}$ in the null hypothesis $r=0$, is $\delta r = 0.0037$,
and is limited by sample variance of the lensing signal.  This sets
the detection limit of gravitational waves from a measurement of
$B$-mode polarization with Clover.

\section{Conclusions}
We have briefly described the scientific aims and design of a new
instrument, Clover, which is designed to measure the $B$-mode
polarization of the CMB. It is intended that deployment of the
experiment to the site will be phased over three years and the full
instrument consisting of three 256-horn arrays operating at 90, 150
and 220~GHz is planned to be operational by 2008.  It is expected that
with a two-year observing schedule, Clover will allow measurements of
the $B$-mode polarization over the $\ell$-range 20--1000, with a
detection limit of $r\sim0.005$.

\end{document}